\documentclass[12pt]{iopart}
\usepackage{iopams,bm}

\usepackage{ifpdf}
\ifpdf
    \usepackage[pdftex]{graphicx}
    \usepackage{color}
    \usepackage[pdftex,
               breaklinks=true,
               colorlinks=true,
               ]{hyperref}
\else
    \usepackage[dvips]{graphicx}
    \usepackage[hypertex,breaklinks=true,colorlinks=false]{hyperref}
\fi

\begin{document}
\title[Detecting spontaneous symmetry breaking in finite-size spectra]
{Detecting spontaneous symmetry breaking in finite-size spectra
of frustrated quantum antiferromagnets}
\author{Gr\'egoire Misguich}
\address{Service de Physique Th\'eorique, URA CNRS 2306, CEA Saclay\\
91191 Gif sur Yvette, France}
\author{Philippe Sindzingre}
\address{Laboratoire de Physique Th\'eorique de la Mati\`ere
condens\'ee, Univ. P. et M. Curie\\
75252 Paris Cedex, France}
\date{}

\begin{abstract}
Exact diagonalization is  a  powerful numerical technique  to  analyze
static    and dynamical   quantities  in   quantum  many-body  lattice
models. It  provides unbiased information  concerning quantum numbers,
energies and wave-functions  of the low-energy eigenstates for  almost
any kind of microscopic  model.   The information about  energies  and
quantum numbers is particularly  useful to detect possible spontaneous
symmetry breaking at $T=0$.  We review some of the advances
in the field  of frustrated quantum  magnets which have been  possible
thanks to    detailed  symmetry  analysis of   exact  diagonalizations
spectra. New results concerning the kagome and star lattice Heisenberg
antiferromagnets are presented.
\end{abstract}
\submitto{\JPCM}

\section{Introduction}

There exists a wide range of numerical techniques to deal with quantum
many-body problems. In the field of lattice spin Hamiltonians, quantum
Monte Carlo, density matrix renormalization group (DMRG) are among the
most powerful, in particular   because they allow to  study accurately
very large  systems. None of them is  however efficient  to study {\em
frustrated} models in $D>1$.  These  systems, in $D=2$ in  particular,
can present  a large  variety of phases,  with potentially  new exotic
phases of matter~\cite{ml05}.

Although limited to  small system  sizes, exact diagonalizations  (ED)
give very  valuable piece of information  on these frustrated systems.
It can provide (almost) any physical  quantity: energies, gap, quantum
numbers,  static and dynamical   correlations, thermodynamics, ... The
method has {\em no bias}, and requires no  particular knowledge on the
low-energy physics.   It  has  been  applied   from the beginning   of
numerical  investigations of  quantum  spin models.\footnote{In  1964,
Bonner  and   Fisher~\cite{bf64}  studied      a  by   ED      11-site
spin-$\frac{1}{2}$ chain.}

The first question  to  address in  order to  characterize a state  of
matter, concerns  the (spontaneously) broken symmetry(ies).  A natural
way to   do   so,  is to   evaluate  possible  order  parameters  from
appropriate correlation functions  in systems of increasing sizes and
study if they might remain finite in  the thermodynamic limit.  Thanks
to the rapid growth of the computing power, the available system sizes
have  increased   a  lot   since  the  first  ED  studies.\footnote{In
Ref.~\cite{ldlst05} the  ground-state of  a 40-site spin-$\frac{1}{2}$
model  was  computed.  A  star  lattice with  42  sites was studied in
Ref.~\cite{star}.}  Still, these sizes (a  few tens of  sites at most)
are sometimes too small to decide if  a given order parameter vanishes
in the infinite size limit.  In these cases an  approach based on {\em
spectrum analysis} can be very useful to detect broken symmetries.  It
is simply based on the fact that a spontaneous symmetry breaking (SSB)
at $T=0$  implies some ground-state (quasi-)degeneracy  on finite-size
spectra.  This paper illustrates the   power of this method through  a
few examples   and  references  to  recent  studies    of  2D  quantum
antiferromagnets.  Some new  data on the spin-$\frac{1}{2}$ Heisenberg
model on  the kagome and  ``expanded-kagome'' lattices (EKL) will also
be presented.

\section{Basics of the exact diagonalization method}

This section is a brief presentation of  the ED method. It is intended
for non  specialist   and is  not specific  to   studies of frustrated
magnets.  Some   readers  may therefore    wish  to  go   directly  to
Sec.~\ref{sec:ssb}.

\subsection{Lanczos  method}

The  Lanczos method  is  a   numerical iterative algorithm   specially
efficient to find a few  extremal eigenvalues and eigenstates (at both
ends  of the spectrum) for  a big sparse   hermitian matrix.  Starting
from an initial random  vector, one iteratively builds an  orthonormal
basis  of the Hilbert space  in which the Hamiltonian is tri-diagonal.
The  spectrum  of   the growing   tri-diagonal   matrix  progressively
converges to the spectrum of the Hamiltonian.  The power of the method
comes  from the   fact that  the  {\em  extremal} eigenvalues converge
first.  In particular,   the lowest  eigenvalue  of  the  tri-diagonal
matrix   converges  (exponentially  fast) to    the  true ground-state
energy. The number  of iterations required  to obtain the ground-state
with  a  good accuracy  is {\it  much} smaller than  the Hilbert space
dimension.  In a typical case, the 2 first eigenvalues  of a matrix of
size  $10^6$  are obtained  with machine  accuracy   in about only 100
iterations (matrix-vector multiplications).  Once the ground-state has
been found, a variant of the Lanczos  algorithm may be used to compute
dynamical correlations (for a recent example, see Ref.~\cite{lp04}) or
the spin-stiffness~\cite{es95}.

\subsection{Other algorithms - Full diagonalization}
Standard  algorithms/libraries (like LAPACK) allow  to obtain the {\it
full}  spectrum of an hermitian  matrix.   The available Hilbert space
dimensions are then however much smaller (typically a few $10^4$ for a
few  gigabytes of memory)   than   with  the  Lanczos  method.    Full
diagonalization  allow   to   compute exactly  the  finite-temperature
properties of    the system.  This may   also  be a step  to   solve a
generalized diagonalization  eigenvalue  problem  as arises  when  the
microscopic spin model has  been projected onto a variational subspace
in  a non-orthogonal basis,   such as the  subspace of  first-neighbor
valence  bond   coverings   (see   Refs.  \cite{ze95,mm00}   and
paragraph~\ref{ssec:ek} below for an example).

\subsection{Symmetries}

The  ED method   becomes really powerful  only when   the spectrum  is
computed separately in each   symmetry sector.  The symmetry  analysis
can be    described  mathematically on general  grounds.     One first
determines  the symmetry group    $\mathcal   G$ of  the   Hamiltonian
$\hat\mathcal{H}$\footnote{In the case of an Heisenberg  model on a 2D
finite lattice with  periodic boundary conditions,  this  group is the
direct  product  of $SU(2)$  (global  spin rotations) and  the lattice
symmetry group.  The latter  usually contains translations as  well as
point group symmetries (lattice rotations and axis reflections).}  and
the               irreducible          representations            (IR)
$\gamma_0,\gamma_1,\cdots,\gamma_n$   of $\mathcal {G}$. The projector
$\hat\Pi_\alpha$  onto  the  subspace which  transforms  according  to
$\gamma_\alpha$         (symmetry        sector     $\alpha$)    reads
$\hat\Pi_\alpha=\frac{1}{|\mathcal{G}|}\sum_{g\in\mathcal{G}}\chi_\alpha(g)\hat
g^{-1}$  where $\chi_\alpha(g)=\Tr[\gamma_\alpha(g)]$ is the character
of   $\gamma_\alpha$ for the  group  element $g$  (noted $\hat g$ when
acting on physical states). By construction, the $\hat \Pi_\alpha$ are
orthogonal  and  commute   with  $\hat\mathcal{H}$.    Thus  one   can
diagonalize $\hat\mathcal{H}$   separately   in  the  image    of each
projector  (symmetry sectors).  In  fact,  one can further refine  the
decomposition  of the Hilbert  space.   For  a given sector  $\alpha$,
$\hat\Pi_\alpha$     can      be     written       as       a     sum:
$\hat\Pi_\alpha=\sum_{i=1}^{d_\alpha}\hat\pi^i_\alpha$.  $d_\alpha$ is
the  dimension of the  IR $\gamma_\alpha$,  the $\hat\pi^i_\alpha$ are
defined                                                             by
$\hat\pi^i_\alpha=\frac{1}{|\mathcal{G}|}\sum_{g\in\mathcal{G}}\gamma_\alpha^{(i,i)}(g)\hat
g^{-1}$  and  $\gamma_\alpha^{(i,i)}(g)$ is the $i^{\rm  th}$ diagonal
element of  $\gamma_\alpha(g)$  (in  some    arbitrary basis).     The
$\hat\pi^i_\alpha$ also   form   a family   of  orthogonal  projectors
commuting with  $\hat\mathcal{H}$.  The   spectrum needs not    to  be
computed for   all  subspaces  ${\rm   Im}[\hat\pi_\alpha^{i}]$.   One
spectrum (e.g.   $i=1$) will be enough since,  in sector $\alpha$, all
the $\hat\pi_\alpha^{i=1,\cdots,d_\alpha}$ give the same eigenvalues (hence a
degeneracy equal to the IR dimension $d_\alpha$).

Such a symmetry analysis   offers many advantages.   The decomposition
into stable  orthogonal subspaces allows to  work with smaller vectors
and  requires less memory and CPU  time.  The density  of states being
much smaller in each sector than in the  full spectrum, convergence is
reached faster.  In the subspace ${\rm Im}[\hat\pi_\alpha^{i=1}]$, the
eigenvalues  are  (generically)   {\em non-degenerate}.   The   actual
degeneracy of an energy level (hard to find with Lanczos if symmetries
where not used) is simply given  by the IR  dimension $d_\alpha$.  The
knowledge of the IR  of each energy level is  very important to detect
SSB.

In many cases  the  IR  of the  full  lattice  symmetry group  can  be
obtained from  {\em one-dimensional} representations of some subgroups
(the IR   are {\it  induced}  by some   subgroup representations).  In
practice, this amounts first to fix  the lattice momentum $k$ and then
look for IR of  its {\it little  group} (which leaves  $k$ invariant).
Such  a    simplified   description  in   terms   of   one-dimensional
representations of a   subgroup   is  however not   always    possible
(examples: symmetry group of the icosahedron~\cite{pierre}.)

\subsubsection{Implementations }

Usually, a full projection  in individual symmetry sectors of  $SU(2)$
with total spin $S$ is not implemented.  This requires large extra CPU
time/memory  and is not necessary.   Instead, one merely  use a $U(1)$
subgroup, that  is conservation  of  $S^z_{\rm total}$ with  spin-flip
symmetry in the   sector  $S^z_{\rm total}=0$  which  project  on  the
subspaces with $S$  even or odd.   For $SU(2)$ invariant Hamiltonians,
an  energy  level  of spin $S$  will  then   be degenerate in  sectors
$|S^z|\le   S$ which allows   to infer the  value  of $S$.  As regards
lattice symmetries, a fully automatic treatment can be implemented. As
an example, the spectra shown in Sec.~\ref{ssec:ek} were obtained by a
program  with  automatic     detection  of lattice   symmetries    and
construction  of  the   corresponding  IR  (using GAP\cite{gap})   and
application of  the   appropriate projectors $\pi$.    However, to our
knowledge, no software with such  an systematic/automatic treatment of
lattice symmetries is publicly available at present.

\section{Detecting spontaneous symmetry breaking}
\label{sec:ssb}
\subsection{Quantum numbers}
\label{ssec:qn}
The existence of a SSB (at $T=0$) in the thermodynamic limit, has
direct consequences on the structure of the low-energy spectrum
(degeneracies, quantum numbers).  Although the SSB only takes place
in the thermodynamic limit, this structure is often visible on
(very) small systems, provided the that the model is not too  close
to a (quantum) phase transition.  It is  a very useful signature of
SSB in ED
studies.

The basic idea is the following.  Let $|\psi\rangle$ be a ground-state
of the system and $\mathcal{G}$ the symmetry group of the Hamiltonian.
If $|\psi\rangle$  is a {\it broken symmetry  state}, there exists, by
definition, at least one group element $g\in
\mathcal{G}$ under         which          $|\psi\rangle$  is not
invariant: $|\langle\psi|\hat g|\psi\rangle|<1$.   The linear space
$V_{gs}$ generated by all the states $\{\hat g |\psi\rangle \;,\;g\in
\mathcal{G}\}$ has thus a dimension  $d>1$, it defines a
(non-trivial)  linear   representation   $\Gamma$   of  $\mathcal{G}$.
Because any  $g\in\mathcal{G}$ commutes with  the Hamiltonian, all the
states of $V_{gs}$ are degenerate ground-states.  The decomposition of
$\Gamma$ onto IR $\gamma_\alpha$ of $\mathcal{G}$: $\Gamma=\bigoplus_\alpha
n_\alpha \gamma_\alpha$, is obtainable from  group theory.  One  finds
that the  spectrum contains exactly  $n_\alpha$ ground-state(s) in the
symmetry sector labeled by the IR $\alpha$ (for $n_\alpha\ne 0$).  The
multiplicities $n_\alpha$ may be obtained from the following character
representation formula:
\begin{equation}
n_\alpha=\frac{1}{|\mathcal{G}|}\sum_{g\in\mathcal{G}}
\chi_\alpha(g^{-1})\sum_{|i\rangle} \langle i|\hat g|i\rangle
\end{equation}
where the  states  $|i\rangle$    form an orthonormal  basis   of  the
ground-state  manifold  $V_{gs}$.   This   can  also   be  written  as
$n_\alpha=\sum_i ||\hat\Pi_\alpha|i\rangle||^2$, so  that $n_\alpha>0$
if  and only if a broken  symmetry state has  a non-zero projection on
sector $\alpha$. Of course, these  multiplicities {\em only} depend on
the symmetry properties of $V_{gs}$. They can therefore be computed by
choosing   a   simpler  state,  $|\psi_0\rangle$,    (without  quantum
fluctuations  for instance) belonging  to the same  ``phase''. At this
stage  the problem  of finding the  $n_\alpha$ has  been reduced  to a
purely  group-theoretical  problem.\footnote{  This   is  particularly
simple for one-dimensional  representations.   In that case  one first
determines the subgroup $H$  of  symmetry operations which  leaves all
the broken  symmetry states   invariant.     Only the  sectors   where
$\chi_\alpha(h)=1$ for {\it all} $h\in H$ can have $n_\alpha>0$. }
If the system has a finite
size, the different  sectors with $n_\alpha>0$ will  not  be exactly
degenerate. Still,  for  a  large enough system  they  should become
the lowest eigenstates of the spectrum.
\subsection{Discrete broken symmetry, valence-bond crystals}

For    a discrete broken  symmetry,   there   are  a finite number  of
degenerate ground-state in the thermodynamic limit, separated by a gap
(finite  in the thermodynamic limit)  from physical excitations.  This
(quasi-)degeneracy  does   not depend  on   the system  size (provided
boundary  conditions  do not  frustrate   the order).   In  frustrated
magnets, valence-bond       crystal   (VBC)     are    the    simplest
examples~\cite{ml05}. The procedure described in~\ref{ssec:qn} gives a
straightforward relation  between the  VBC   pattern and  the  quantum
numbers of  the  quasi-degenerate ground-states.   For  a large enough
system, one can therefore directly read off  the spatial symmetries of
the possible VBC from the spectrum,  without computing any correlation
function.   To  confirm  the  SSB in   the  thermodynamic  limit,  one
eventually has to check the exponential decay  of the energy splitting
between the  quasi-degenerate  ground-states.  A complementary  method
based  on reduced  density  matrices~\cite{fmo06} can then  be used to
determine ``automatically'' the order parameter and the VBC pattern.

Recent ED studies have exhibited  various VBC phases. Some 2D examples
for $SU(2)$ symmetric   spin-$\frac{1}{2}$ models are  on   the on the
honeycomb~\cite{fsl01}                   and                    square
lattices~\cite{lws02,fmsl03,ldlst05,mlpm06}.  The  case of  the kagome
antiferromagnet will   be  discussed  from  a  VBC   point of  view in
Sec.~\ref{ssec:kag}      and  a new      example   on   the  EKL    in
Sec.~\ref{ssec:ek}.  VBC orders may occur  in more complex models, but
ED remain there limited to quite small sizes.  For example, ED studies
of a 2D $SU(4)$ spin-orbital model  on the square lattice indicate the
possibility of a VBC phase there~\cite{bzm02}

\subsection{Broken continuous symmetry: N\'eel and nematic  phases}

The structure  of the spectrum is richer   for {\it continuous} broken
symmetries  than for discrete ones.   First,  the existence of gapless
Goldstone  modes makes the  distinction  between ``ground-states'' and
``excitations'' more subtle.    And since the   ground-state  subspace
($V_{gs}$) has  a infinite dimension in  the  thermodynamic limit (one
can perform infinitesimal rotations  of  the initial  broken  symmetry
state), the number of quasi-degenerate ground-states  has to grow as a
function of the system size.

Frustrated  magnets  exhibit a remarkably large variety of phases with
broken   continuous   symmetries.  In a    N\'eel   phase, the lattice
spontaneously breaks up in sublattices in which the all spins point in
the same  direction (up to    quantum zero-point fluctuations).    The
simplest  example  is the two-sublattice  collinear structure realized
(at $T=0$) in the (unfrustrated) antiferromagnetic Heisenberg model on
the square  lattice.  But  frustration can lead  to  much more complex
structures          with           non-collinear~\cite{bernu,ldlst05},
non-coplanar~\cite{dslp05}          sublattice         magnetizations,
order-by-disorder effects~\cite{lblp95}, nematic phases with    broken
$SU(2)$ symmetry but no sublattice magnetization~\cite{ldlst05,sms06},
etc.

To obtain  the quantum  numbers  of  the finite-size ground-states  in
N\'eel  systems, one has to decompose  a  broken-symmetry N\'eel state
$|\psi_0\rangle$  (with  a well-defined   direction for the sublattice
magnetizations) onto  the  IR  of  the  $SU(2)$  and lattice  symmetry
groups.   As explained above,  we can consider  a ``classical'' N\'eel
state with maximum  sublattice magnetization (each sublattice is fully
polarized).  For  the  two-sublattice problem $|\psi_0\rangle$ can  be
chosen            as               an            Ising           state
$|\psi_0\rangle=|\uparrow\downarrow\uparrow\downarrow\cdots\rangle$. Although
trivial,      the    two-site            case    is       instructing:
$|\psi_0\rangle=|\uparrow\downarrow\rangle\sim
|S=0\rangle+|S=1,S^z=0\rangle$ is the linear  combination of a singlet
state  {\it   and}  a  triplet ($S=1$)  state.    Similarly, a general
classical N\'eel state for $N$ (even) sites will have a finite overlap
on {\it all total  spin sectors}  from  $S=0$ to $S=N/2$.   Still, the
weight of the different spin  sectors decreases with increasing $S$ in
a N\'eel state (which has  no net magnetization  in any direction) and
one       has     in      particular     $\langle\psi_0|\vec    S_{\rm
total}^2=S(S+1)|\psi_0\rangle\sim N$.  From this we can guess that, in
fact, only spin sectors from $S=0$  to $S\sim\sqrt{N}$ are required to
construct    a  state with     finite  (but  not  maximum)  sublattice
magnetization   and    non-zero  quantum   fluctuations.     A  simple
semi-classical argument allows to guess the finite-size scaling of the
energies         for       these    states     (so-called   ``Anderson
tower''~\cite{anderson},        or                 QDJS             in
Refs.~\cite{bernu,lhuillier05}). As for classical spins, one expects a
finite  uniform  susceptibility per site $\chi$   in  a quantum N\'eel
phase.  The total  magnetization   $M$ is therefore given  by  $\simeq
N\chi B $ in presence of an (infinitesimal) applied field $B$.  $M$ is
also obtained   by  minimizing $E(M)-M\cdot B$,  where   $E(M)$ is the
energy   of   the  lowest  state  with   total   spin  $S=M$  (in zero
field). Combining the two leads to $E(S)=S^2/(2N\chi)$.  This spectrum
of  the  Anderson tower corresponds  physically  to the (slow) quantum
dynamics   of the ``rigid  body''  made by the  macroscopic ($\sim N$)
sublattice  magnetizations, it should  not  be confused with  physical
spin-wave excitations   (higher in the   spectrum).  As  expected, the
states with $S\lesssim \sqrt{N}$ collapse onto the ground-state (up to
a total energy  $\mathcal{O}(1)$) in the $N\to\infty$  limit so that a
broken symmetry  state (some combinations of  the QDJS)  has an energy
per site $\epsilon\sim 1/N\to   0$.  This structure  (including  space
group     quantum  numbers,   not    discussed  here)    involves many
($\sim\sqrt{N}$)  states and imposes   strong  constraints on the  low
energy spectrum.  It has been observed numerically in finite-size
spectra   for    a   large    number   of   quantum   antiferromagnets
(\cite{lhuillier05} and Refs. therein).  This tower structure offers a
very efficient  way   to recognize  systems with    continuous  broken
symmetries, as it  already shows up on very  small systems (the energy
gaps in  the Anderson tower ($\sim 1/N$)  decay faster than the finite
size corrections to the sublattice magnetization ($\sim 1/\sqrt{N}$)).
Collinear versus  non-collinear   N\'eel states  can also  be  readily
discriminated  from there spectra.   A collinear structure has a tower
with  exactly {\it  one}  level per  value of  $S$ 
whereas the tower of a non-collinear system
(with $\geq  3$ sublattices) contains $2S+1$ levels  in the total spin
sector $S$.   We also  mention that tower   structures have also  been
discussed (and observed numerically)  for larger symmetry  group, like
$SU(4)$, in the context of spin-orbital models~\cite{pmfm03}.

A similar analysis can be  carried out  for quantum  $p$ (resp.   $n$)
spin-nematics~\cite{nematic},  which  are  systems with broken $SU(2)$
symmetry but no sublattice magnetization. A simple picture is that the
spins spontaneously select a preferred  oriented (resp.  non-oriented)
{\it plane}, but no particular {\it direction} (the order parameter is
an  anti-symmetric  (resp.   symmetric) rank   2   tensor in the  spin
variables).  Again,  such  a  broken  symmetry    state is a    linear
combination  of many finite-size   eigenstates  of the systems,   with
specific (and  predictable  by  group theory) total  spin  and spatial
quantum numbers.  As examples, we  mention two recent ED studies which
exhibited the tower structures  associated to a 2D $p-$~\cite{ldlst05}
and $n-$~\cite{sms06} nematics.

\subsection{$J_e-J_t$ model on the expanded kagome lattice}
\label{ssec:ek}

The    EKL (also    dubbed    star
lattice~\cite{star}) is obtained from the  kagome lattice by splitting
each site into  two, and inserting an ``expanded'' bond connecting
the  two neighboring  triangles.   Its  plaquettes  are triangles  and
dodecagons and  all the  sites are  equivalent.
Concerning frustration,
the  classical antiferromagnetic Heisenberg model on the EKL
has a huge degeneracy equivalent to that on the kagome lattice~\cite{star}.
Since  the   EKL has  two  kinds  of  bonds,
(triangle  and  ``expanded'' bonds),    it  is natural to  consider  a
$J_t-J_e$ spin-$\frac{1}{2}$  Heisenberg model (both antiferromagnetic
here). To our knowledge this model has only been studied for $J_e=J_t$
so far~\cite{star}.

The limit $J_e\gg J_t$ is simple: the system is made of weakly coupled
{\it  bonds},   the ground-state is   unique   (spin singlet) and  all
excitations are gapped ($\Delta\sim J_e$), it is an {\it explicit VBC}
(terminology  of  Ref.~\cite{ml05}) without  any  broken symmetry. The
model has been shown~\cite{star} to    be in   this      phase  at least     up to
$J_t/J_e=1$.

The  limit of weakly coupled triangles ($J_e\ll J_t$) is more
interesting as the  ground-state is extensively degenerate at  $J_e=0$
(4 degenerate ground-states per triangle).  As  in Mila's work
on the   trimerized  kagome lattice~\cite{mila98},   the   degenerate
perturbation theory    for  $J_e/J_t\ll 1$ can    be  formulated as a
spin-chirality Hamiltonian. In fact, Mila's self-consistent mean-field
solution of this Hamiltonian almost directly applies on the EKL.
As a result, one finds an   {\it extensive} number of (mean-field)
ground-states    which  are in   one-to-one  correspondence  with some
particular singlet coverings   of the EKL  (dubbed ``super coverings''
hereafter).  These  singlet coverings  are  those which  maximize  the
number of occupied  $J_t$ bonds (as expected  since $J_t\gg J_e$). But the
extensive degeneracy is clearly an  artifact of the mean-field  approximation.

\begin{figure}
\includegraphics[width=7.7cm]{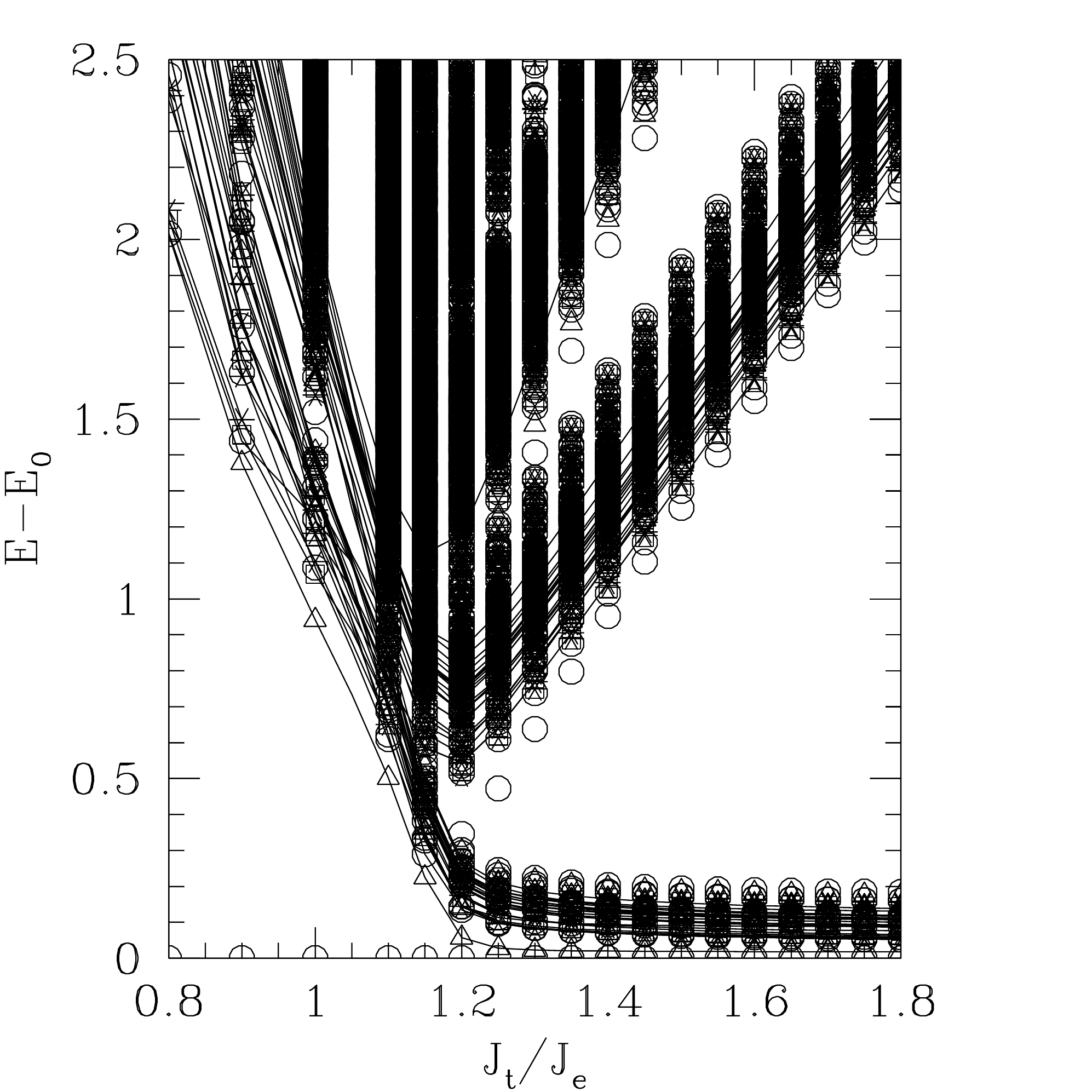}\hspace*{-1.3cm}
\includegraphics[width=7.7cm]{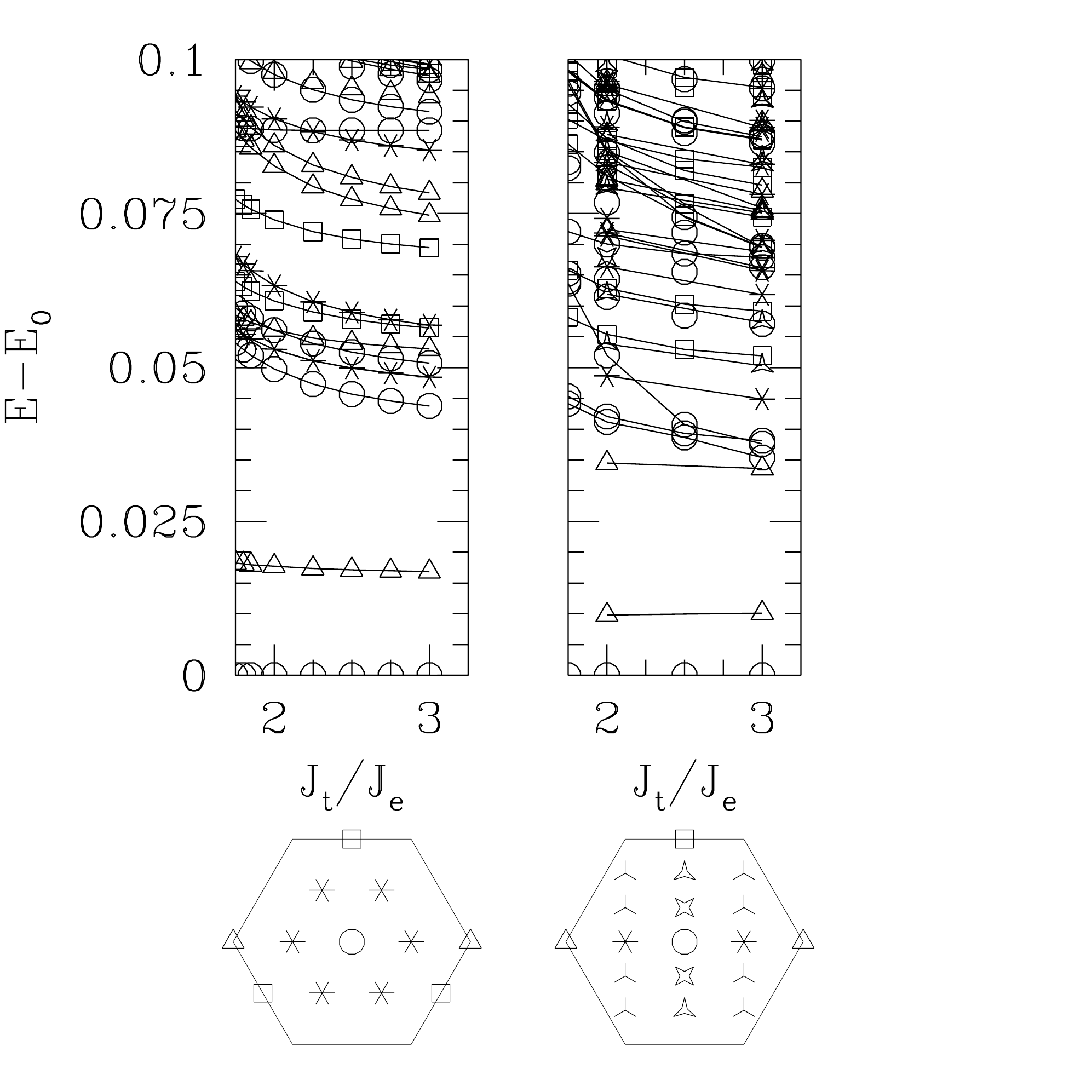}\hspace*{-1.7cm}
\includegraphics[width=2.8cm]{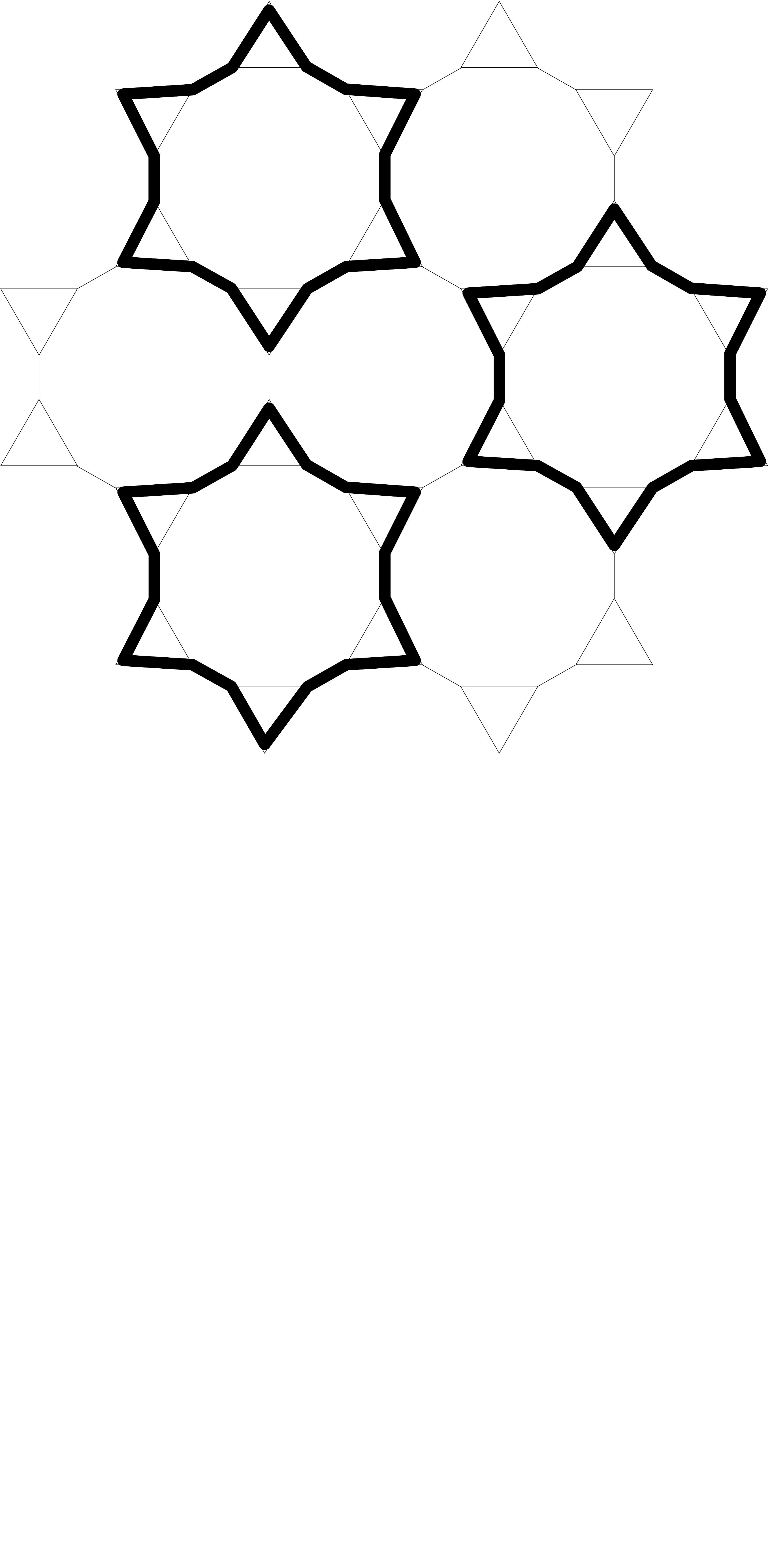}
\caption{Spectrum of the $J_e-J_t$ spin-$\frac{1}{2}$ Heisenberg
Hamiltonian   (on   the EKL)  projected     in the  first-neighbor RVB
subspace.  Left and  middle   spectra: lattice with  12  dodecagons (72
sites).   Right spectrum:  18  dodecagons (108  sites).   The  wave-vectors symbols
are indicated below, in the corresponding  Brillouin zones.  The
VBC pattern proposed for  $J_t\gtrsim 1.3 J_e$  is shown on the right of the figure
(the singlet ``stars'', with 18 sites, are marked with fat bonds).
}
\label{fig:ek}
\end{figure}

To  go beyond, we  performed some ED of  the EKL Heisenberg model {\it
restricted the subspace  of first-neighbor valence  bond coverings} (a
method initiated for the   kagome antiferromagnet~\cite{ze95}).  As  a
justification, we note that   this variational RVB   subspace contains
both the exact $J_t=0$ ground-state (a singlet on  each $J_e$ bond) as
well as all the mean-field solutions  (super coverings) of $J_e/J_t\ll
1$.  The   subspace dimension is  $2^{(N+1)}$   for $N$ dodecagons and
periodic   boundary     conditions\footnote{As  for    the      kagome
lattice~\cite{ze95,msp02}, there is  a mapping between dimer coverings
and Ising pseudo-spin  configurations.}, which  is much  smaller  than
total Hilbert space  dimension  $2^{6N}$.   Because different  singlet
coverings are not  orthogonal, the matrices are not  sparse and we had
to resort to full diagonalizations.  Using all lattice symmetries, the
complete spectrum has been obtained for $N=12$ and  $16$ and a part of
the spectrum for $N=18$\footnote{The  two 4-dimensional IR's (subspace
dimensions $> 29.10^3$) corresponding  to the 8 wave-vectors indicated
by 3-leg symbols  in the Brillouin  zone (right of  Fig.~\ref{fig:ek})
could {\it  not} be computed for  $N=18$ (108 sites) because of memory
limitations.}.   The results  are   summarized Fig.~\ref{fig:ek}.  The
spectrum evolves from a regime  with one ground-state  and a large gap
($J_t/J_e\lesssim 1.1$)  to  a  regime  with  many  low-energy  states
($J_t/J_e\gtrsim 1.3$).   As expected, the  number of  such low-energy
states matches  the number  of  super coverings.  For  $J_t/J_e\gtrsim
1.3$ one  energy level,  corresponding to  an  IR  of dimension 2  and
momentum in the zone corner,  is significantly below the other excited
states (and goes  further down  from  $N=12$ [middle spectrum] to  $N=18$
[right spectrum]).  This is what  one expects for an  SSB toward a 3-fold
degenerate VBC.  From the  momenta ($k=(0,0)$ and $k=(0,\pm 4\pi/3)$),
such VBC  should be  invariant under any  3-step  translations and the
simplest guess is a crystal with  resonating ``stars'' (analog to that
found in  the hexagonal lattice  quantum dimer model~\cite{msc01}), as
shown at the right of Fig.~\ref{fig:ek}.  This is also natural because
the energy gained by singlet resonances is  larger of small loops, and
the 18-site  stars  are  the shortest  loops  made  out  (three) super
dimers.   Although  further investigations   are  certainly  needed to
confirm the  existence of this VBC,  the spectra  (in this variational
approximation) clearly show that it is the most likely scenario.

\subsection{Antiferromagnetic spin-$\frac{1}{2}$ Heisenberg
model on the kagome lattice}
\label{ssec:kag}

It has been argued that the kagome antiferromagnet could form a VBC at
{\it very}  low  energies.   One  scenario~\cite{ms91,ze95,ns03} is  a
crystal  (VBC-1  in the following),   which maximizes  the  number  of
``perfect''  hexagons in a  way  which preserves the $2\pi/3$ rotation
symmetry  of the lattice  (see Fig.~2 of Ref.~\cite{ns03}).  Using the
method    of Sec.~\ref{ssec:qn},   we    computed the quantum  numbers
associated to this crystal for a  36-site sample\footnote{ The quantum
numbers $\sigma$ and $R_2$ may depend on the relative  sign of the two
singlet   coverings   participating in   a  plaquette resonance (here,
hexagons  and stars).  These signs  were  determined by minimizing the
energy (by  ED) in a variational  a subspace of first-neighbor singlet
coverings.}.     The  lowest   exact    eigenstates     (obtained   by
ED~\cite{waldtmann}) matching these quantum numbers are marked by * in
the column VBC-1  of Table~\ref{tab:k36}.  Although one cannot exclude
a deep reorganization  of  the low-energy  levels  when increasing the
system size, these states are {\it not} the lowest ones.  Similarly, a
4-fold degenerate  crystal of resonating 12-site ``stars''\cite{sma02}
would  require the two levels   indicated in the  column ``VBC-2''  to
become the lowest states.  The quantum numbers of the ``columnar'' VBC
proposed in Ref.~\cite{ba04}  (degeneracy  24) correspond to  the last
column (VBC-3).   In  all cases,  the crystallization would  require a
complete reshuffling of the low-energy spectrum.

\begin{table}
\begin{center}
\small
\begin{tabular}{|c|c|c|ccc|c|c|c|c|}
\hline
Number & $2\langle \vec S_i\cdot \vec S_j\rangle$ &
$k$ & $R_3$ & $R_2$ & $\sigma$ & Deg. & VBC-1 & VBC-2	& VBC-3		\\ 
\hline
 1	& -.43837653  &  $0$ &  1       	&1  & 1 &  1 &*b  &*	&*\\	
 2      & -.43809562  &  $B$ &$e^{\pm 2i\pi/3}$	&   &   &  4 &  &	& \\	
 3      & -.43807091  &  $0$ &$e^{\pm 2i\pi/3}$	&1  &   &  2 &  & 	&*\\	
 4      & -.43799346  &  $0$ &  1       	&1  & 1 &  1 &  & 	& \\	
 5      & -.43785105  &  $C$ &          	&   & 1 &  6 &*b & 	& \\	
 6      & -.43758510  &  $0$ &  1       	&-1 & 1 &  1 &  & 	&*\\	
 7      & -.43758455  &  $A$ &          	&1  & 1 &  3 &*b &	&*\\	
 8      & -.43751941  &  $C$ &          	&   &-1 &  6 &*a & 	& \\	
 9      & -.43721566  &  $0$ &  1       	& 1 &-1 &  1 &*a & 	&*\\	
 10     & -.43718796  &  $0$ &$e^{\pm 2i\pi/3}$	&1  &   &  2 &  & 	&*\\	
 11     & -.43714765  &  $A$ &          	&-1 &-1 &  3 &  & 	&*\\	
 12     & -.43705108  &  $0$ &$e^{\pm 2i\pi/3}$	&-1 &   &  2 &  & 	&*\\	
 13     & -.43703981  &  $B$ &  1       	&   &1  &  2 &*b & 	& \\	
 14     & -.43703469  &  $A$ &          	&-1 & 1 &  3 &  & 	&*\\	
 15     & -.43685867  &  $0$ &  1       	&-1 &-1 &  1 &  & 	&*\\	
 16     & -.43685319  &  $B$ &  1       	&   &-1 &  2 &*a  & 	& \\	
 17     & -.43683757  &  $A$ &          	& 1 &-1 &  3 &*a &* 	&*\\	
\hline
$\cdots$ & $\cdots$ & $\cdots$ &  & $\cdots$ &  & $\cdots$ & & & \\
\hline
44 & -0.43474519      &  $0$ &$e^{\pm 2i\pi/3}$	&-1 &   &  2 &  & 	&*\\	
\hline
\end{tabular}
\end{center}

\caption{
Low energy levels of the Heisenberg  model on a 36-site kagome lattice
(all are total spin singlets).  Momentum  $k=A$ refers to the 3 middle
points of the (hexagonal) zone boundary, $B$ to the two corners of the
Brillouin zone.  The  6  points $C$ correspond  to  $\pm  B/2$.  $R_3$
(resp.  $R_2$) indicates (when  applicable) the phase factor  acquired
by the wave-functions under a $2\pi/3$  (resp.  $\pi$) rotation around
the center  of an  hexagon.  $\sigma$ is  the parity  under reflection
about the momentum direction.  For $k=0$, the reflection axis $\sigma$
is chosen  parallel to a  lattice bond.  The eigenstates which quantum
numbers match those of VBC ground-states (see  text) are marked by $*$
in the last columns.  There are in fact two  kinds of VBC-1, depending
on the symmetry of the  resonating stars (or  pin-wheels): {\it even} or {\it odd}
under axis reflections. The  12 levels marked  with  $a$ refer to  the
``odd-star'' VBC-1, and those with $b$ to the ``even-star'' VBC-1.  }

\label{tab:k36}
\end{table}

\ack

We  are grateful to A. L\"auchli,  C.~Lhuillier and L.~Pierre for many
discussions and collaborations on these topics and  to J.  Schulenburg
for correspondence about the software Spinpack.

\end{document}